# Megahertz FDML Laser with up to 143nm Sweep Range for Ultrahigh Resolution OCT at 1050nm

Jan Philip Kolb[1,2], Thomas Klein[2,3], Mattias Eibl[1,2], Tom Pfeiffer[1,2], Wolfgang Wieser[2,3] and Robert Huber[1]

[1]Institut für Biomedizinische Optik, Universität zu Lübeck, Lübeck, Peter-Monnik-Weg 4, 23562 Lubeck, Germany

[2]Lehrstuhl für BioMolekulare Optik, Fakultät für Physik, Ludwig-Maximilians-Universität München, Oettingenstr. 67, 80538 Munich, Germany

[3]Optores GmbH, Gollierstr. 70, 80339 Munich, Germany

## ABSTRACT

We present a new design of a Fourier Domain Mode Locked laser (FDML laser), which provides a new record in sweep range at ~1µm center wavelength: At the fundamental sweep rate of 2x417 kHz we reach 143nm bandwidth and 120nm with 4x buffering at 1.67MHz sweep rate. The latter configuration of our system is characterized: The FWHM of the point spread function (PSF) of a mirror is 5.6µm (in tissue). Human in vivo retinal imaging is performed with the MHz laser showing more details in vascular structures. Here we could measure an axial resolution of 6.0µm by determining the FWHM of specular reflex in the image. Additionally, challenges related to such a high sweep bandwidth such as water absorption are investigated.

**Keywords:** Optical coherence tomography, OCT, tunable laser, Fourier domain mode locking, FDML, MHz OCT

## 1. INTRODUCTION

Optical coherence tomography (OCT) enables the non-invasive study of morphological features of the human retina with high resolution. The transverse resolution is determined by the optical design of the system and aberrations of the human eye. If the latter is corrected with adaptive optics individual photoreceptors can be visualized [1] – even at MHz A-scan rate [2]. The axial resolution is independent of the optics and scales with the square of the central wavelength of the light source and with the inverse of its bandwidth. Since non-sweeping light sources such as super continuum lasers have considerably higher bandwidth than swept laser sources [3-5], spectral domain OCT is still superior over swept source OCT regarding axial resolution as shown in Fig. 1. With spectral domain OCT, ophthalmic imaging with axial resolutions down to ~2µm has been demonstrated [6]. While this might not be required for the diagnosis of most diseases, the diagnosis of some pathologies where individual cell layers are affected might benefit [7].

The purpose of this research is to push the axial resolution of swept source OCT. Here, we achieve axial resolutions of 5-7µm comparable to typical clinical systems but with a MHz-OCT system. In a first step, we increased the bandwidth of our swept laser source – the FDML laser. Recent technical advances of semiconductor optical amplifiers (SOAs) and chirped fiber Bragg gratings (cFBGs) enabled us to double the sweep range from 72nm to up to 143nm. Then we investigate challenges related to such high sweep bandwidths.

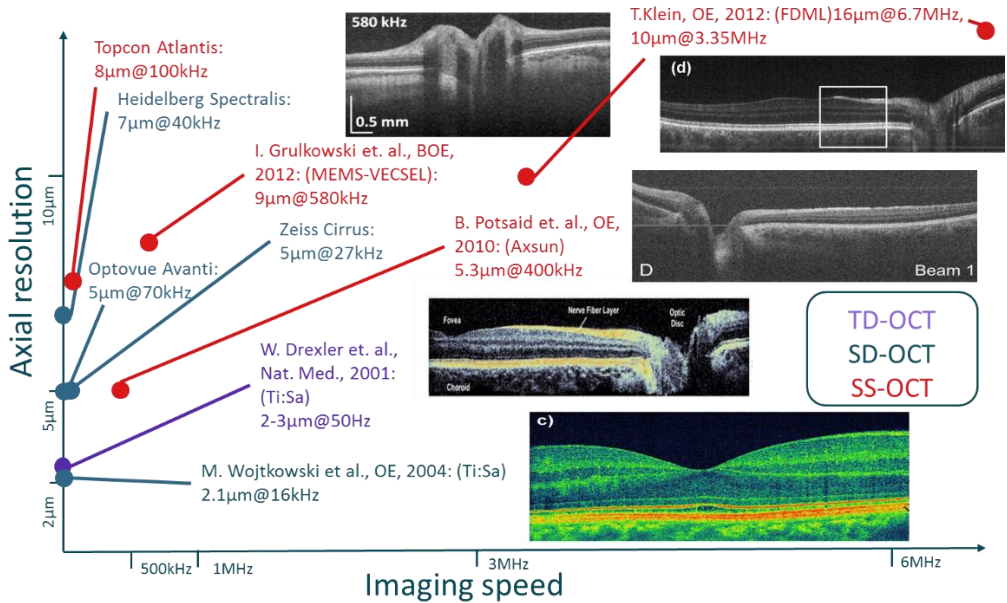

Figure 1: Overview of imaging speed vs. axial resolution for selected OCT systems. The axial resolution is determined either by measuring the FWHM of the PSF of a mirror [6, 8-10] or a retinal structure[7].

## 2. METHODS

Figure 2 displays the schematics of our FDML-Laser [11]. FDML lasers exhibit the unique feature being capable of wavelength sweep repetition rates well into the MHz range [9, 12-18], because they do not suffer from inherent physical constraints with respect to wavelength sweep rate [19-23]. Hence, they found widespread applications from ultrafast OCT [24-28], over non-destructive sensing and testing [29-33] to optical molecular [34, 35] and functional imaging [36, 37] and even short laser pulse generation [38]. The extension of the accessible wavelength range of FDML lasers to the 1060nm spectral region [12, 17, 39-42] enabled high quality ultra-fast retinal swept source OCT imaging with good penetration into the choroid [9, 12, 17]. Compared to our previous version [9], we improved the gain bandwidth of our laser gain medium by 40nm through incorporating a newly developed SOA from Innolume (SOA-1020-110-HI-27dB). Together with a homebuilt fiber Fabry-Perot filter achieving a free spectral range of over 180nm at a hardware resonance frequency of 417 kHz, we could tune over a spectral bandwidth of 143nm. To make use of this broader sweep range and still have low noise and stable operation of our FDML laser, we had to compensate the intra-cavity dispersion for this increased spectral bandwidth. We achieved this by two measures. Firstly, by designing and implementing cFBGs (manufactured by Teraxion) to remove the main mismatch in roundtrip time for different wavelengths. And secondly, by minimizing the remaining dispersion with an optimized combination of fiber types with different dispersion characteristics.

Figure 3 shows the extended sweep ranges of 143nm and 120nm at 417kHz and 1.67MHz, respectively. We apply a buffering scheme for the 1.67 MHz laser where the current of the SOA gain medium has to be modulated rapidly. We assume that the lower sweep range with buffering is due to a too low modulation bandwidth of the used laser diode controller (Wieserlabs WL-LDC10D).

The point spread function (PSF) was measured with the reference and recalibration arm of our Michelson interferometer at 1.67MHz sweep rate and 120nm bandwidth. Data acquisition was performed with an Alazar Tech ATS9360 digitizer card at 1.8GS/s and a sampling depth of 12bit. This results in an imaging range of 2.1mm. For in vivo imaging in humans, we used our standard MHz imaging setup as presented in [43]. The scanning protocol included 1900 A-scans per B-scan and a B-scan density of 1 B-scan/6µm. The latter matches the axial resolution of 7.5µm (FWHM) quite well.

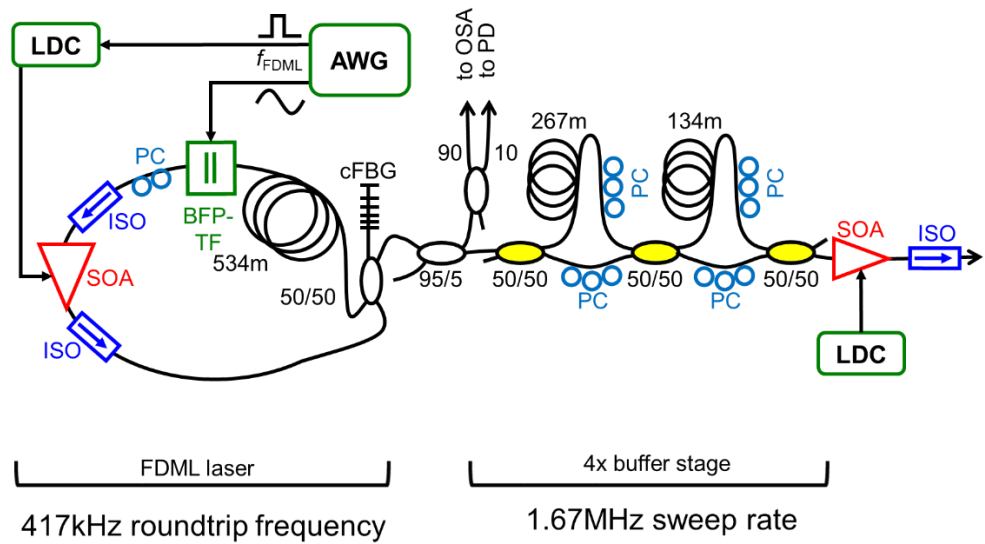

Figure 2: Schematic of our FDML laser and optional buffer stage: AWG: Arbitrary waveform generator, LCD: Laser diode controller, ISO: optical isolator, PC: polarization controller, BFP-TF: Fabry-Pérot filter, OSA: Optical spectrum analyzer, PD: photo diode.

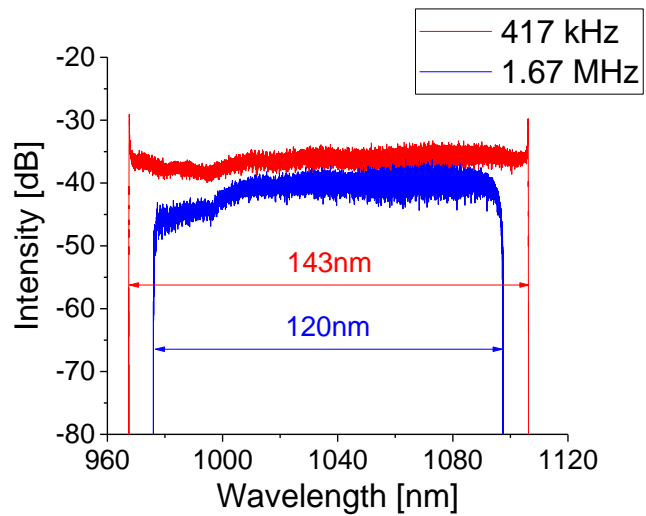

Figure 3: Spectra of the new FDML laser at 417kHz and at 1.67MHz sweep rate.

# 3. RESULTS AND DISCUSSION

Figure 4 shows the results of the PSF measurement. The PSF has a FWHM of 7.4µm (in air) which corresponds to 5.6µm in tissue with a refractive index of ~1.3.

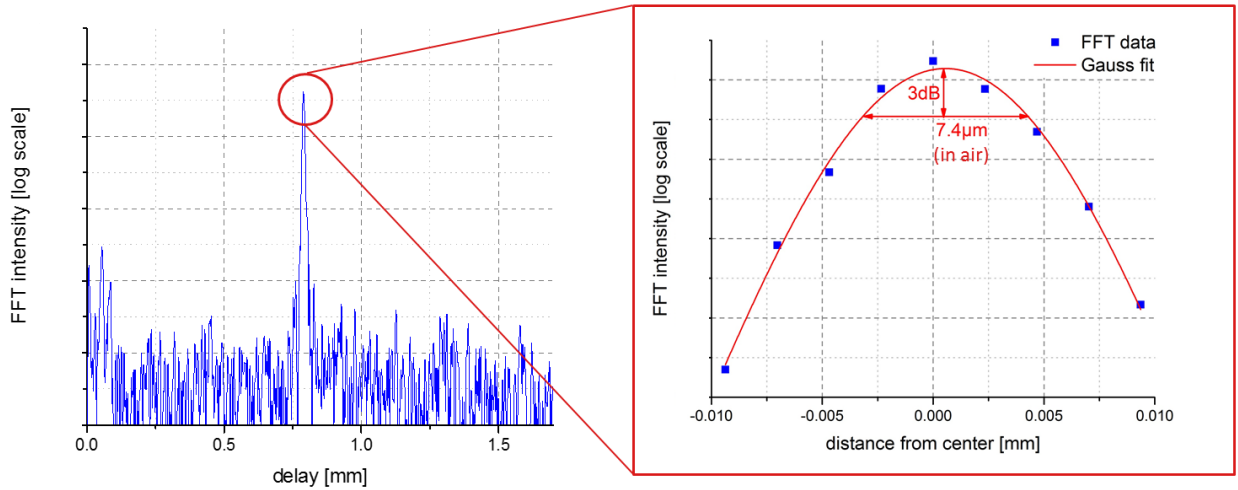

Figure 4: PSF of a mirror at 800µm imaging depth. The FWHM is 7.4µm in air. This corresponds to an axial resolution of 5.6µm in tissue.

A healthy volunteer was imaged to test the imaging capabilities of the new FDML design. We used an incident power of 1.6mW. In order to compare the new results with our previous lower bandwidth design, we imaged at 120nm sweep bandwidth and at 70nm. The results are displayed in Figure 5. Although, the images look quite similar at a first glance, a higher level of detail in the choricapillaris and fine vessels in the upper layers is obvious in the magnified images. The FWHM of a specular reflex was 6.0µm, which corresponds very well with the PSF measured from a mirror.

For a more detailed investigation of the axial resolution- which is determined by the width and shape of the spectrum of our swept source laser-we measured the output spectrum at various positions in our setup with an optical spectrum analyzer: Directly at the FDML laser, after the booster SOA and at the photodetector of the interferometer after passing 4cm of water. The results are displayed in Figure 6. Next, we Fourier transformed the spectra in order to get the theoretically possible axial resolution by measuring the FWHM of each FFT. With the spectrum directly from the FDML laser, we get a FWHM of 4.9µm (in tissue, as for all following), after the light is amplified by the booster SOA a FWHM 5.3µm was determined. In order to simulate imaging of the human retina, we added a 2cm cuvette of water to our sample arm and measured the spectrum directly at the photodetector of our interferometer. The FFT showed a possible resolution of 5.9µm, which is very close the 6.0µm measured directly in the image. As shown in Figure 7, the water absorption in the eye varies strongly over the whole sweep range, which explains the change in spectrum and reduction of axial resolution [44].

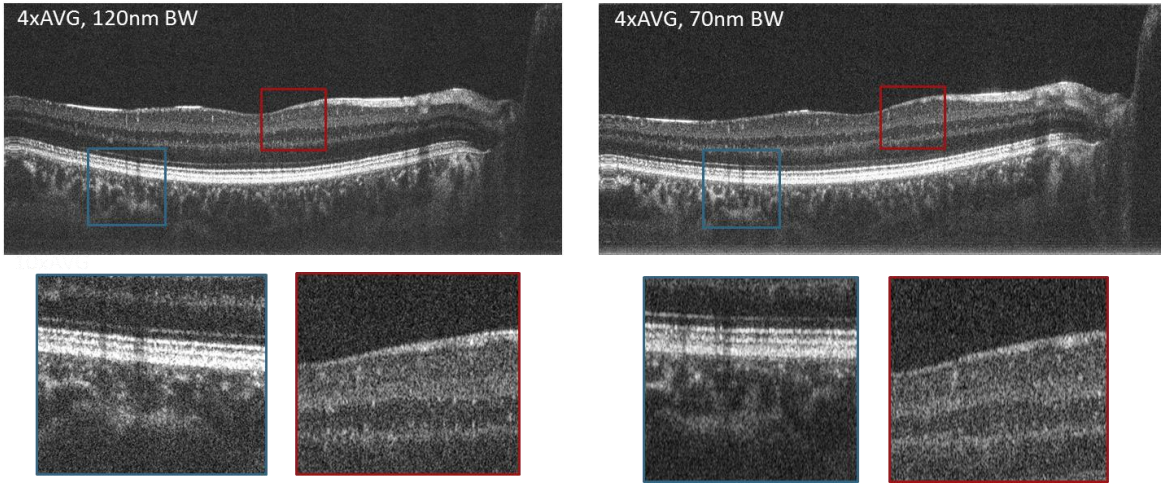

Figure 5: 4 times averaged B-Scan of a heathy volunteer at 120nm and at 70nm sweep bandwidth. Magnifications of the boxed sections are shown below.

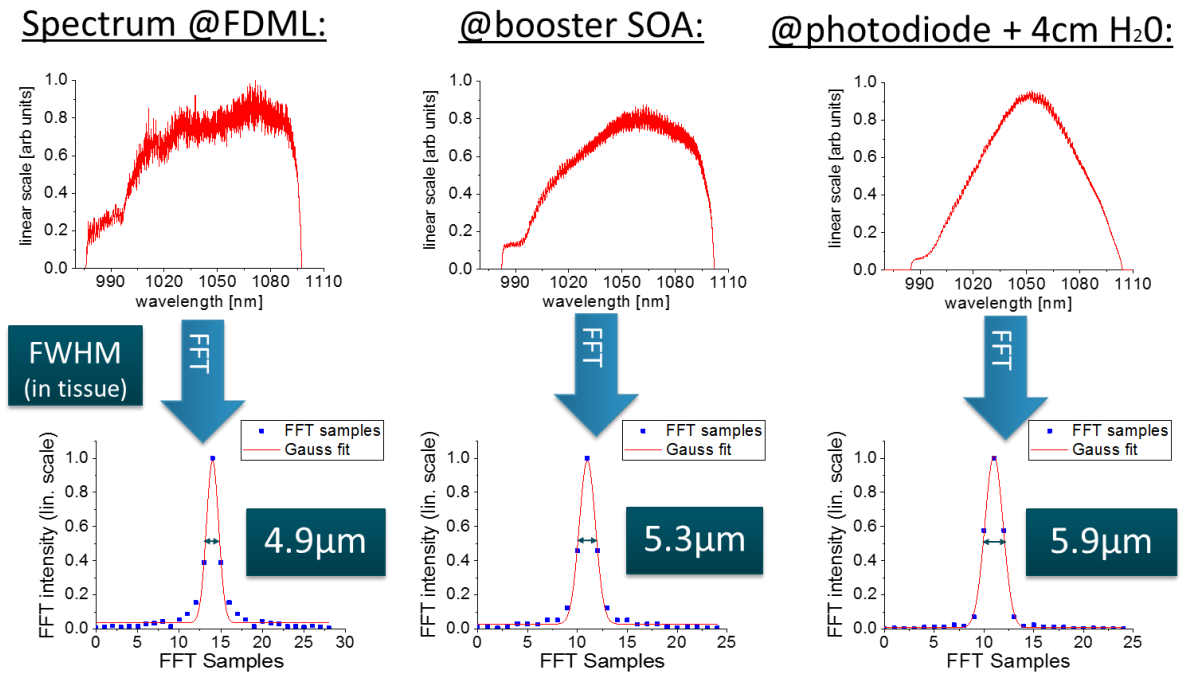

Figure 6 top row: Spectra measured directly at the FDML laser, after the booster SOA and at the photodetector of the interferometer after passing 4cm of water. Bottom row: FFT transformation of each spectrum with FWHM in tissue.

Considering the resolution values we achieved only resolution levels comparable to standard commercial spectral domain OCT, so right now the system has no "ultra-high resolution", yet. Considering the FFT of a Gaussian shaped spectrum with 120nm at 1050nm center wavelength the axial resolution is 3µm. However, as can be seen the 3dB bandwidth of our spectrum is much smaller, especially considering the effective spectrum after water absorption. Therefore, we believe that there is still potential for improving the axial resolution by spectral shaping by modulating the current of the SOA [45, 46] or an optical filter [47].

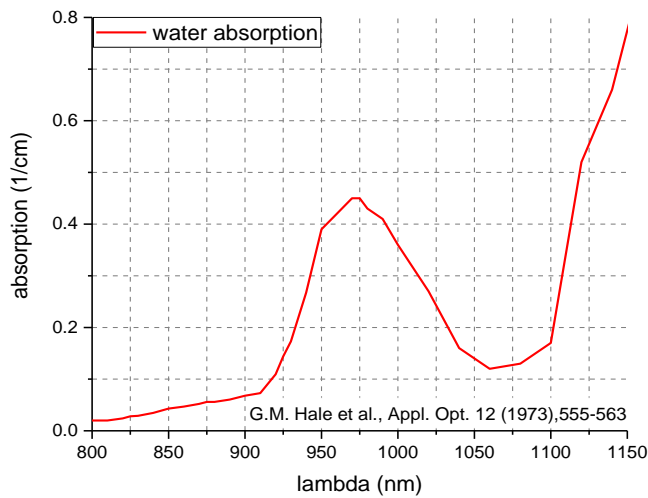

Figure 7: Absorption of water in the near infrared [48].

## 4. CONCLUSION AND OUTLOOK

We demonstrated new record high bandwidths of 143nm and 120nm at 417 kHz and 1.67 MHz A-scan rate of a swept laser source at ~1µm center wavelength. The benefits of the higher bandwidth were shown in in vivo human retinal OCT imaging, where the visibility of vascular structures was enhanced. An axial resolution of 6.0µm could be measured by determining the FWHM a specular reflex in an in vivo image. This is within the range of typical resolutions for commercial OCT systems, but at MHz speed. Improvements regarding the spectral shape should improve axial resolution and therefore image quality even further. Moreover, we plan to use a laser diode controller with a higher modulation bandwidth in order to achieve a higher sweep bandwidth for the buffered operation. Our results indicate that swept source retinal MHz-OCT with ultra-high resolution is feasible and provides detailed and rapidly acquired OCT images for better diagnosis